\documentstyle[12pt,twoside]{article}
\def\greaterthansquiggle{\raise.3ex\hbox{$>$\kern-.75em\lower1ex\hbox{$\sim$}}}
\def\lessthansquiggle{\raise.3ex\hbox{$<$\kern-.75em\lower1ex\hbox{$\sim$}}}
\newcommand{\beq}{\begin{equation}}
\newcommand{\eeq}{\end{equation}}
\newcommand{\beqa}{\begin{eqnarray}}
\newcommand{\eeqa}{\end{eqnarray}}
\newcommand{\beqan}{\begin{eqnarray*}}
\newcommand{\eeqan}{\end{eqnarray*}}
\newcommand{\ba}{\begin{array}}
\newcommand{\ea}{\end{array}}
\newcommand{\no}{\nonumber}

\newcommand{\ra}{\rightarrow}

\newcommand{\ve}{\varepsilon}

\newcommand{\wt}{\widetilde}

\newcommand{\R}{{\cal R}}

\def\nz{\ifmmode {I\hskip -3pt N} \else {\hbox {$I\hskip -3pt N$}}\fi}
\def\zz{\ifmmode {Z\hskip -4.8pt Z} \else
       {\hbox {$Z\hskip -4.8pt Z$}}\fi}
\def\qz{\ifmmode {Q\hskip -5.0pt\vrule height6.0pt depth 0pt
       \hskip 6pt} \else {\hbox
       {$Q\hskip -5.0pt\vrule height6.0pt depth 0pt\hskip 6pt$}}\fi}
\def\rz{\ifmmode {I\hskip -3pt R} \else {\hbox {$I\hskip -3pt R$}}\fi}
\def\cz{\ifmmode {C\hskip -4.8pt\vrule height5.8pt\hskip 6.3pt} \else
       {\hbox {$C\hskip -4.8pt\vrule height5.8pt\hskip 6.3pt$}}\fi}

\def\au{{\setbox0=\hbox{\lower1.36775ex%
\hbox{''}\kern-.05em}\dp0=.36775ex\hskip0pt\box0}}
\def\ao{{}\kern-.10em\hbox{``}}

\normalsize
\begin{document}
\bibliographystyle{plain}
\begin{titlepage}
\begin{flushright}
UWThPh-2000-35\\
\today
\end{flushright}
\vspace{2.5cm}
\begin{center}
{\Large \bf On unbounded bodies with finite mass: \\ 
asymptotic behaviour}\\[50pt]
R. Beig*   \\
Institut f\"ur Theoretische Physik \\ Universit\"at Wien\\
Boltzmanngasse 5, A-1090 Wien \\
and \\
M. Karadi \\
KFKI Research Institute for Particle and Nuclear Physics \\
H-1525 Budapest, P.O.B. 49, Hungary

\vfill 
{\bf Abstract}
\end{center}

There is introduced a class of barotropic equations of state (EOS) which become 
polytropic of index $n = 5$ at low pressure. One then studies asymptotically 
flat solutions of the static Einstein equations coupled to perfect fluids
having such
an EOS. It is shown that such solutions, in the same manner as the vacuum
ones, are conformally smooth or analytic at infinity, when the EOS is
smooth
or analytic, respectively.

\vfill
\noindent
*) Supported by ``Fonds zur F\"orderung der wissenschaftlichen Forschung'',
Projekt Nr. P12626--PHY.
\end{titlepage}

In this Letter we introduce and study asymptotically flat solutions of the
static Einstein equations for a source consisting of perfect-fluid matter
with a specific class of equations of state (E'sOS). These E'sOS are chosen 
in such a way that there exist solutions where the support of the
energy-momentum tensor is unbounded. In other words, there are solutions
for which the fluid extends to infinity but becomes dilute at infinity
fast enough so that the gravitational field is asymptotically flat, in
particular the total mass is finite.

Before introducing the EOS we recall the field equations for a static
metric on ${\bf R} \times M$ of the form
\beq
ds^2 = g_{\mu\nu} dx^\mu dx^\nu = - e^{2U} dt^2 + e^{-2U} \gamma_{ij}
dx^i dx^j,
\eeq
where $U,\gamma_{ij}$ depend only on the spatial coordinates $x^i$.
For the source we have that
\beq
T_{\mu\nu} = \rho u_\mu u_\nu + p(g_{\mu\nu} + u_\mu u_\nu),
\eeq
where $u_\mu u_\nu g^{\mu\nu} = - 1$ and $u_\mu = f \xi_\mu$ with
$\xi^\mu \partial_\mu = \partial/\partial t$. The Einstein equations
\beq
G_{\mu\nu} = 8\pi \; T_{\mu\nu}
\eeq
give rise to
\beq
\Delta U = 4 \pi (\rho + 3p)e^{-2U}
\eeq
\beq
\R_{ij} = 2D_i U D_jU - 16 \pi \gamma_{ij} p e^{-2U},
\eeq
where $\Delta$ and $\R_{ij}$ are respectively the Laplacian and the Ricci
tensor of $\gamma_{ij}$. From the contracted Bianchi identities for
$\gamma_{ij}$ we obtain the condition for hydrostatic equilibrium in the
form
\beq
D_i p = - (\rho + p) D_i U .
\eeq
If we have a barotropic EOS given by $\rho = \rho(p)$ and $\rho$ and $p$
are positive, both $\rho$ and $p$ can be viewed as functions of $U$. Then,
from Eq. (6),
\beq
\frac{dp}{dU} = - (\rho + p).
\eeq
Instead of prescribing $\rho(p)$ we can thus give an EOS in the parametric
form
\beq
\rho = \rho(U), \qquad p = p(U).
\eeq
A specific example, and the one which originally motivated the present work,
is the so-called Buchdahl EOS \cite{Bu} which can be written as
\beq
\rho = \rho_0(1 - e^U)^5, \qquad 
p = \frac{1}{6} \rho_0 e^{-U}(1 - e^U)^6, \qquad
0 < \rho_0 = \mbox{const},
\eeq
which is equivalent to
\beq
p = \frac{1}{6} \frac{\rho^{6/5}}{(\rho_0 - \rho)^{1/5}}, \qquad
0 < \rho < \rho_0.
\eeq
It is known \cite{Bu} that, for each value of the central pressure, there
exists a spherically symmetric solution of (4,5) which is asymptotically
flat with matter extending to infinity (``Buchdahl solution''). In the
paper \cite{Be1} it was shown that any asymptotically flat solution of
(4,5) with
the EOS (10) coincides with a Buchdahl solution (in particular: is
spherically symmetric). However in \cite{Be1} it was assumed that, like 
in the vacuum case, asymptotic flatness at infinity implies conformal
smoothness at infinity \cite{Be2}. This gap was closed in the 
paper \cite{WS}. In the present work we consider
the following, much more general, class of E'sOS:
\beq
(\rho + 3p)e^{-2U} = U^5 \phi(U^2)
\eeq
\beq
p e^{-2U} = U^6 \psi(U^2).
\eeq
Here $\psi : {\bf R}^+ \ra {\bf R}$ is a smooth or analytic function and
$\phi$ is chosen so that (7) is satisfied, i.e.
\beq
\phi(x) = - 2 [3 \psi(x) + x \psi'(x)].
\eeq
One checks that (9) is a special case of (11, 12, 13) with $\psi$ analytic.
(Explicitly we have that $\psi(x) = \frac{\rho_0}{6} (\sinh \frac{\sqrt{x}}{2}
/ \frac{\sqrt{x}}{2})^6$.)
In the case where the mass is positive we will have that $U$ is strictly 
negative at least near infinity. Choosing, then, $\psi(0) > 0$, both
$\rho$ and $p$ are positive and $\rho \sim 6(\psi(0))^{-1/6} p^{5/6}$ as
$p \ra 0$. Thus our E'sOS behave asymptotically like that of a polytrope of
index 5.

In the following we assume that $M$ is diffeomorphic to ${\bf R} \setminus
{\bf B}_R(0)$ where ${\bf B}_R(0)$ is the closed ball of Euclidean radius
$R$ centered at the origin. The pair $(U,\gamma_{ij})$ is required to
satisfy the decay conditions
\beq
U = O^\infty(1/r), \qquad
\gamma_{ij} - \delta_{ij} = h_{ij} = O^\infty(1/r),
\eeq
where $r^2 = x^i x^j \delta_{ij}$ and $O^\infty(F(r))$ means that the
quantity in question is $O(F)$, its derivative is $O(F'(r))$, a.s.o.
The Buchdahl solution satisfies these criteria. We will, in this paper,
not study the general existence question of solutions satisfying the
asymptotic conditions. The expectation is that, for any EOS subject to
(11, 12, 13), there exist solutions having, say, finitely many but
arbitrary multipole moments. (In the vacuum case this statement is also
still a conjecture.) The situation with respect to global solutions, i.e.
ones where $M \cong {\bf R}^3$ is different since one expects -- and in
many cases knows \cite{Be3,Li} -- that these have to be spherically
symmetric and thus, for given EOS, are determined by a single parameter
such as the central pressure. On the other hand, restricting to spherical
symmetry from the outset and given an EOS satisfying (11, 12, 13) and such
that $\rho$ and $p$ are positive: it is not clear whether solutions for
some central pressure have finite radius, are not even asymptotically flat,
or are at the borderline between these cases, namely of infinite radius
but finite mass, i.e. have the asymptotic behaviour studied in this paper.
(Strictly speaking, in the present context, we should look at the quantity
$U$ on the r.h. sides of (11, 12) as {\bf some} parameter, not necessarily
the
gravitational potential, since, e.g. for finite radius, $U$ does not go
to zero at the surface of the body if it is required to vanish at infinity.)

Returning to the topic of this paper, namely asymptotic solutions, our main
result is as follows:
\paragraph{Theorem:} Let $(U,\gamma_{ij})$ be a solution of (4, 5) with EOS
satisfying (11, 12, 13) on ${\bf R}^3 \setminus {\bf B}_R(0)$ with decay
conditions (14) and positive mass. Then it is, in a sense in detail spelled
out later, conformally smooth or analytic, depending on whether $\psi(x)$
is smooth or analytic.

The proof follows the pattern of that for the vacuum case \cite{Si,Be2},
so we shall merely outline it. We first derive an explicit expression for
$(U,\gamma_{ij})$ near infinity up to order $r^{-4}$. To that effect we write
(4, 5) as
\beq
\partial^2 U = h^{ij} U_{,ij} + \Gamma^i U_{,i} + 4\pi U^5 \phi(U^2)
\eeq
with $\Gamma^i = \gamma^{k\ell} \Gamma^i_{k \ell}$ and 
$\partial^2 = \delta_{ij} \partial_i \partial_j$ and
\beqa
\partial^2 h_{ij} &=& 2 \Lambda_{(i,j)} + h^{\ell m} (h_{ij,\ell m} +
h_{\ell m,ij} - h_{kj,i\ell} - h_{i\ell,mj}) + \no \\
&& \mbox{} + 2 \gamma^{mn} \gamma^{p\ell} (\Gamma_{nij} \Gamma_{mp\ell}
- \Gamma_{ni\ell} \Gamma_{mpj}) - 4 U_{,i} U_{,j} - \no \\
&& \mbox{} - 32 \pi \gamma_{ij} U^6 \psi(U^2),
\eeqa
where $\gamma^{ij} = \delta_{ij} - h^{ij}$, 
$\Lambda_i = (h_{ij} - \frac{1}{2} \delta_{ij} h_{\ell\ell})_{ij}$,
$h_{\ell\ell} = \delta_{i\ell} h_{i\ell}$.
The r.h. side of (16) would be $O^\infty(r^{-4})$, if it were not for the
term $2\Lambda_{(i,j)}$. But using a transformation of spatial coordinates
of the form
\beq
\bar x^i = x^i + f^i(x)
\eeq
one may choose $f^i$ with $f^i(x) = O^\infty(\ln r)$ in such a way that,
at least after suitably enlarging the radius $R$ in $M = {\bf R}^3 \setminus
{\bf B}_R(0)$, the quantity $\Lambda_i$ satisfies
\beq
\Lambda_i = O^\infty(( \ln^* r) r^{-3}),
\eeq
where $\ln^* r$ denotes some power of $\ln r$. Thus we now have that
\beq
\partial^2 h_{ij} = O^\infty((\ln^*r)r^{-4}), \qquad
\partial^2 U = O^\infty((\l^* r)(r^{-4}).
\eeq
Using standard (see e.g. \cite{Si}) estimates for the Poisson integral 
we infer
\beqa
U &=& - \frac{M}{r} + O^\infty((\ln^* r) r^{-4}), \\
h_{ij} &=& \frac{t_{ij}}{r} + O^\infty((\ln^* r) r^{-4}),
\eeqa
where $M$ (the mass) and $t_{ij} = t_{(ij)}$ are constants. Using the gauge
condition (18) it now follows that $t_{ij}$ has to vanish. We next insert
(20, 21) into the r.h. sides of (15, 16) and refine the gauge so that
\beq
\Lambda_i = O^\infty((\ln^* r) r^{-4})
\eeq
to obtain
\beq
U = - \frac{M}{r} + \frac{M_i x^i}{r^3} + O^\infty((\ln^*r) r^{-3}),
\eeq
\beq
h_{ij} = M^2 \left( \frac{x_i x_j}{r^4} -- \frac{\delta_{ij}}{r^2}\right)
+ \frac{t_{ijk} x^k}{r^3} + O^\infty((\ln^* r) r^{-3}),
\eeq
where $x_i = \delta_{ij} x^j$ and $t_{ijk} = t_{(ij)k}$ are constants with
$t_{iij} = 0$.

As shown in \cite{Be4} the constants $t_{ijk}$ can be gauged away in a
manner respecting the condition (22). The constant $M_i$ in (23) can also
be disposed of by the translation $\bar x^i = x^i - M_i/M$ (using that
$M$ is positive, in particular nonzero). Inserting the expansions (23, 24)
into the r.h. sides of (15, 16), the above pattern repeats itself. Thus
we obtain
\beqa
U &=& - \frac{M}{r} - \frac{M^3(3cM^2 - 4)}{12r^3} +
\frac{M_{ij} x^i x^j}{r^5} + \no \\
&& \mbox{} + \frac{M_{ijk} x^i x^j x^k}{r^7} + O^\infty((\ln^* r)r^{-5}),
\eeqa
\beqa
h_{ij} &=& \frac{M^2 x_i x_j}{r^4} - \frac{M^2 \delta_{ij}}{r^2} +
\frac{2 M^4 x_i x_j}{3r^6} - \frac{2 M^4 \delta_{ij}}{9r^4} -
\frac{c M^6 x_i x_j}{r^6} - \no \\
&& \mbox{} + \frac{5 M M_{k\ell} x^k x^\ell x_i x_j}{3r^6}
+ \frac{4 M x_{(i} M_{j)k} x^k}{r^6} - \frac{28 M M_{ij}}{27 r^4} - \no \\
&& \mbox{} - \frac{5 M M_{k\ell} x^k x^\ell x_i x_j}{r^8} +
O^\infty((\ln^* r) r^{-5}),
\eeqa
where $c = - 16 \pi \psi(0)$ and the constants $M_{ijk}$ -- the octopole
moment -- satisfy $M_{ijk} = M_{(ijk)}$ and $M_{iij} = 0$.

Next there comes a crucial observation. Let
\beq
\omega = \frac{U^2}{M^2}
\eeq
and define ``unphysical'' coordinates by
\beq
\wt x^i = \frac{x^i}{r^2}.
\eeq
It is then not difficult to check that $\omega$, as a function of
$\wt x^i$, is in the open ball ${\bf B}_{1/R}(0) = \wt M$, four times
continuously differentiable, in fact the fourth derivatives are H\"older
continuous with index $\alpha$ $(0 < \alpha < 1)$, in short:
$\omega \in C^{4,\alpha}$.
(Note that $\omega|_\Lambda = 0$, $D_i \omega|_\Lambda = 0$, where
$\Lambda$ is the point-at-infinity, i.e. $\wt x^i = 0$.)
Furthermore $\wt \gamma_{ij}$ defined by
\beq
\wt \gamma_{ij} = \omega^2 \gamma_{ij}
\eeq
is also $C^{4,\alpha}$ in the coordinate system given by $\wt x^i$.
Armed with this information we can now try to derive field equations for
$(\omega,\wt \gamma_{ij})$ on $\wt M$. We use standard identities on
conformal
rescalings and follow the pattern of \cite{Be2}. Writing, for simplicity,
again $\gamma_{ij}$ for the unphysical metric $\wt \gamma_{ij}$ and
setting $M = 1$ without loss we obtain the equations
\beq
\Delta \omega = 3 \R + \alpha(\omega),
\eeq
\beq
\omega \R_{ij} = \frac{1}{2} (D_i \omega)(D_j\omega) - D_i D_j \omega +
\frac{1}{3} \gamma_{ij} \Delta \omega + \gamma_{ij} \omega \beta(\omega),
\eeq
where
\beqa
\frac{1}{8\pi} \alpha(\omega) &=& \omega \phi(\omega) + 18 \omega \psi(\omega)
+ 12 \phi(\omega), \\
\frac{1}{8\pi} \beta(\omega) &=& - 2 \omega \psi(\omega) - \frac{4}{3}
\phi(\omega).
\eeqa
Contracting (31) and taking the gradient, the vacuum terms without 
prefactor $\omega$ cancel and we obtain
\beq
D_i \R = -(D^j \omega)\R_{ij} + (D_i\omega)(\R - 2\beta + 3\beta') +
(D_i \omega) \omega^{-1} \left( \frac{\alpha}{3} + 3\beta \right).
\eeq
But, luckily,
\beq
\frac{\alpha}{3} + 3 \beta = \frac{1}{3} \omega \phi,
\eeq
so the dangerous last term in (34) is also regular. A similar miracle
occurs after taking $D_k$ of (31) and antisymmetrizing with respect to $j$
and $k$. Using the Ricci identity, the relation
\beq
\R_{ijk\ell} = 2(\gamma_{i[k} \R_{\ell]j} - \gamma_{j[k} \R_{i]\ell}
- \gamma_{i[k} \gamma_{j]\ell}),
\eeq
valid in 3 dimensions and, again, (30, 31) and (34), we finally arrive at
\beqa
\lefteqn{\omega \left[D_{[k} \R_{j]i} - \frac{1}{2}(D_{[k} \omega) 
\R_{j]i}\right] = } \no \\
&& = (D_{[k} \omega) \gamma_{j]i} \left[ -\frac{\alpha}{6} - \frac{\omega\beta}{2}
- 2\beta + 3\beta' + \omega^{-1} \left( \frac{\alpha}{3} + 3\beta\right)
+ (\omega \beta)' + \frac{\alpha'}{3} \right].
\eeqa

Again, by virtue of (32, 33), the dangerous terms drop out. Thus there
results an expression
\beq
D_k \R_{ij} = D_j \R_{ki} + (D_{[k} \omega) \R_{j]i} +
(D_{[k} \omega)\gamma_{j]i} \ve,
\eeq
where $\ve$ is a known smooth or analytic function of $\omega$. Now take
$D^k$ of (38). Using the Ricci and Bianchi identity together with Eq. (36),
and (30, 31) to eliminate second derivatives of $\omega$, we obtain an
equation for $\Delta \R_{ij}$, with a regular r.h. side depending merely
on $\gamma_{ij}$, $\R_{ij}$, $\omega$ and $D_i\omega$. Thus, writing
\beqa
\Delta \omega &=& 3 \R + \alpha \\
\R_{ij} &=& \sigma_{ij} \\
\Delta \sigma_{ij} &=& \ldots
\eeqa
and transforming to harmonic coordinates, the set of equations (39, 40, 41)
becomes an elliptic system for $(\omega,\gamma_{ij},\sigma_{ij})$ with
smooth or analytic coefficients. Consequently $(\omega,\gamma_{ij},
\sigma_{ij})$ which by our previous analysis have been $C^{2,\alpha}$, are
actually smooth or analytic, by virtue of a famous theorem of Morrey
(see Thm 6.7.6 of
\cite{Mo}), and the proof is complete. Let us finally point
out that the
solutions discussed in the present work can be characterized by multipole
moments \cite{Ge} in completely the same way as the vacuum solutions
\cite{Be2}. The details can be found in the diploma thesis by one of us
(M.K.) \cite{Ka}.

\section*{Acknowledgements}
One of us (R.B.) thanks W. Simon for comments on the manuscript and J. 
Thornburg for suggestions on the wording.

\end{document}